\documentclass{webofc}

\usepackage[varg]{txfonts}   
\usepackage{hyperref}
\usepackage{cleveref}
\usepackage{url}
\hypersetup{colorlinks=true,citecolor=blue,urlcolor=blue,linkcolor=blue}
\begin{document}
\title{Precision Measurements of Kinematic Scan for Fluctuations of (Net-)proton Multiplicity Distributions in Au+Au Collisions from RHIC-STAR}
%
%

\author{\firstname{Yige}
\lastname{Huang}\inst{1,2}\fnsep\thanks{\email{yghuang@mails.ccnu.edu.cn}} {(For the STAR Collaboration)}
}

\institute{Key Laboratory of Quark and Lepton Physics (MOE) \& Institute of Particle Physics, Central China Normal University, 430079, Wuhan, China 
\and
GSI Helmholtzzentrum für Schwerionenforschung, 64291, Darmstadt, Germany
          }

\abstract{
This work presents measurements of the rapidity-window dependence of event-by-event net-proton cumulants and proton factorial cumulants in Au+Au collisions at $\sqrt{s_\mathrm{NN}}=$7.7 -- 27 GeV, using high-statistics data from RHIC BES-II.
Protons and antiprotons are identified with improved detector performance within $0.4<p_\mathrm{T}<2.0$ GeV/$c$ and $|y|<0.6$, enabling a wide coverage in momentum space to probe long-range correlations near the QCD critical point.
In the most central collisions, the proton number $\kappa_2/\kappa_1$ and $\kappa_3/\kappa_1$ exhibit power-law scaling with the rapidity window, but with exponents below the theoretical expectation, suggesting that the critical point, if it exists, may lie at higher baryon densities.
A finite-size scaling analysis of the susceptibility and Binder cumulant study points out a critical baryon chemical potential region in 550 -- 650 MeV.
}
\maketitle
\section{Introduction}\label{intro}

One main goal of high-energy heavy-ion collisions is to study the QCD phase diagram and search for a possible critical end point (CEP).
At low baryon chemical potential ($\mu_\mathrm{B} \approx 0$), lattice QCD shows the transition to the Quark Gluon Plasma phase is a smooth crossover~\cite{Aoki:2006we}.
At high $\mu_\mathrm{B}$, some QCD based effective models suggest a first-order phase transition, implying the existence of the critical end point where the transition line ends~\cite{Fodor:2001pe,Bzdak:2019pkr,Pandav:2022xxx}.
However, this has not been confirmed by experiments, and the location of the CEP is still unknown.
The RHIC BES program scans beam energies to explore the QCD phase diagram.
In the second phase of BES, BES-II, STAR upgrades allow detailed and precise studies across a broad $\mu_\mathrm{B}$ range in Au+Au collisions ($\sqrt{s_\mathrm{NN}} =$ 3 -- 27 GeV, corresponding to $\mu_\mathrm{B} = $ 760 -- 156 MeV).

Fluctuations of conserved quantities, such as baryon number, electric charge, and strangeness, are sensitive probes of the critical signal~\cite{Luo:2017faz,Chen:2024aom,STAR:2025zdq}.
In the critical region, the correlation length $\xi$ grows significantly, generating long-range correlations that can be probed through a rapidity scan of (net-)proton number fluctuation.
The long-range pheromone near the critical point underlines the relevance of rapidity scan of fluctuation measures, studied in this proceeding in three aspects: power-law behavior of proton factorial cumulants~\cite{Ling:2015yau}, finite-size scaling with susceptibility in various rapidity windows~\cite{Sorensen:2024mry}, and Binder cumulant as a function of $\mu_\mathrm{B}$~\cite{Sorensen:2024mry}.

\section{Proton Identification}\label{sec:ana_detail}

\begin{figure}[h]
\centering
\sidecaption
\includegraphics[width=7cm,clip]{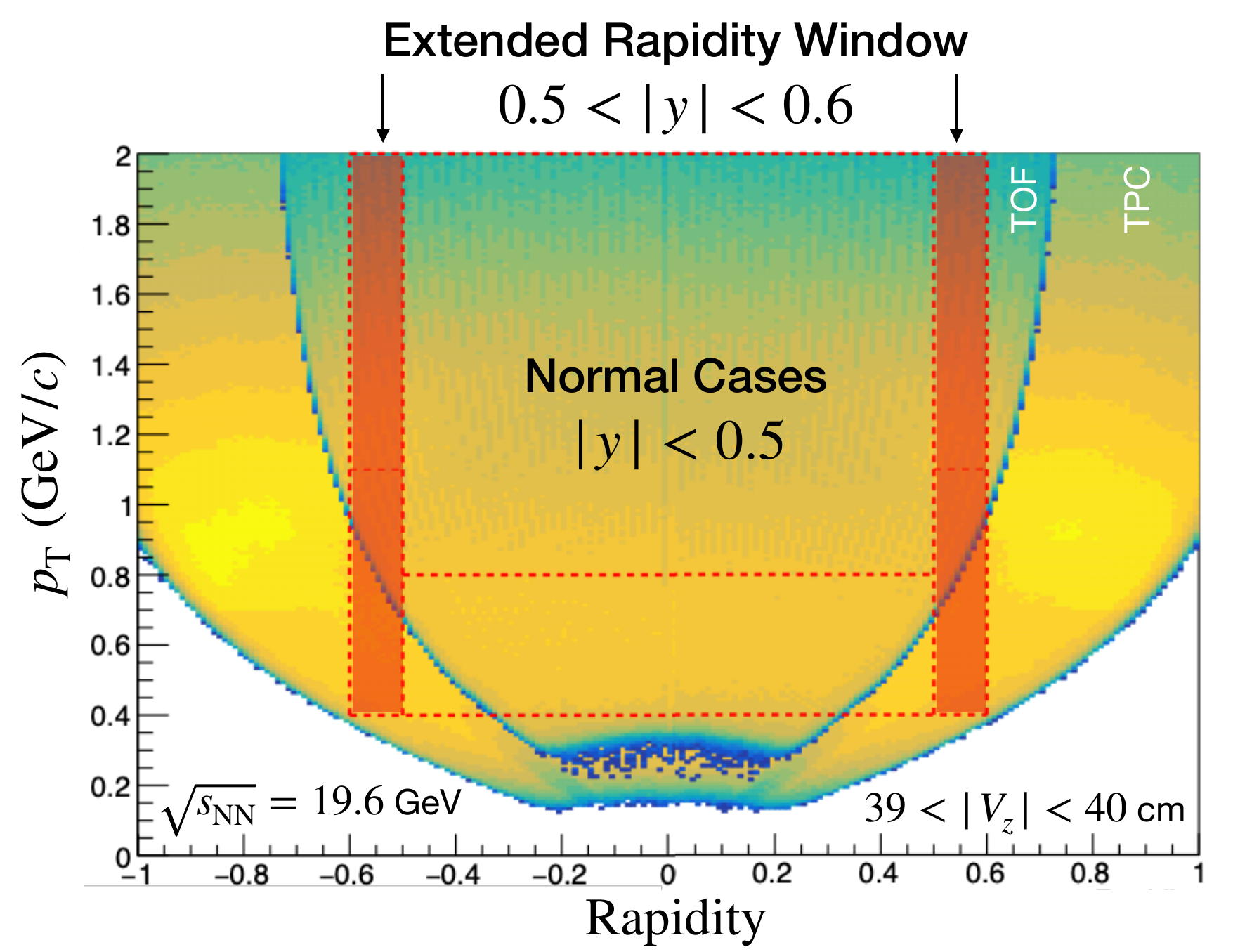}
\caption{The TPC (outer band) and TOF (inner band) acceptance of proton candidates in $y$ -- $p_\mathrm{T}$ plane.
Detector upgrades in BES-II extend the coverage in rapidity up to $|y| < 0.6$ and in transverse momentum of $0.4 < p_\mathrm{T}<2.0$ GeV/$c$.
The protons are identified by energy loss from TPC and mass information from TOF (in high momentum region).
Tracks are taken from Au+Au collisions at $\sqrt{s_\mathrm{NN}}=$ 19.6 GeV, with 39 < $V_z$ < 40 cm for positive rapidity part, and -40 < $V_z$ < -39 cm for negative rapidity part. 
}
\label{fig1:acc}
\end{figure}

The event-by-event proton number is counted for those tracks with good quality and identified by TPC and TOF detectors.
\Cref{fig1:acc} shows the detector acceptance of proton candidates in $y$ -- $p_\mathrm{T}$ plane.
Protons within $|y| < 0.1$ to $|y| < 0.6$, and $0.4 < p_\mathrm{T}<2.0$ GeV/$c$ are counted for fluctuation analysis.
In the analysis, the TPC is the primary detector for proton identification, while the TOF detector is used in the high-momentum region where the TPC alone cannot reliably separate protons from contamination.
The acceptance is incomplete when the primary vertex, where the collision occurs, deviates from the center of detector too far.
The selection of vertex and PID method is optimized to balance proton purity, detection efficiency, and statistics.

\section{Results}\label{sec:res}

\subsection{Proton Factorial Cumulants}

\begin{figure}[h]
\centering
\includegraphics[width=0.75\linewidth,clip]{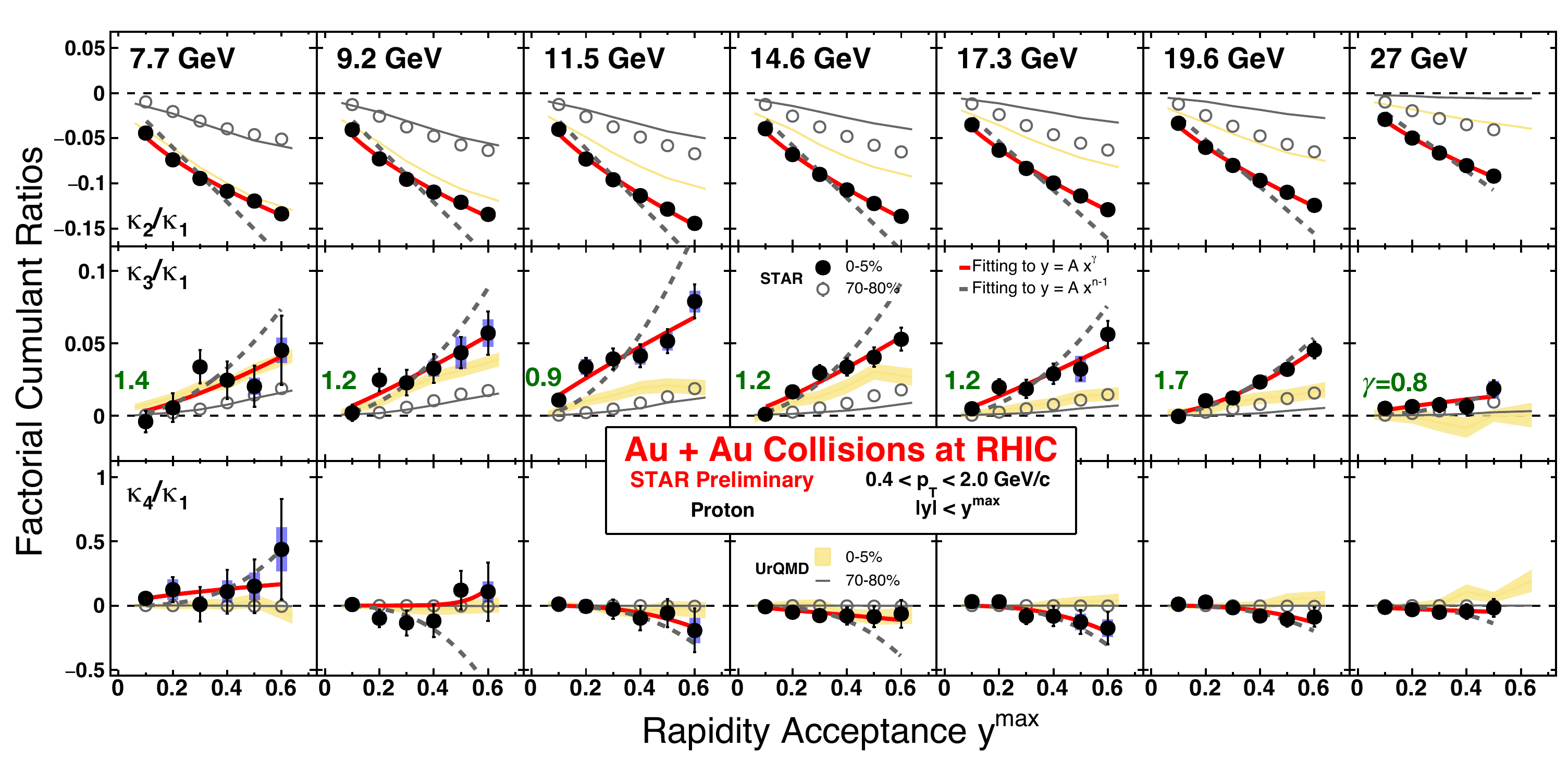}
\caption{Rapidity window size dependence of proton factorial cumulants normalized by $\kappa_1$. Circles denote STAR data: solid for central and open for peripheral collisions. UrQMD results are shown as bands. Red curves indicate power-law fits to solid circles, with third-order exponents labeled in dark green. Gray dashed curves present power function fits with fixed exponent to the expectation~\cite{Ling:2015yau}.
}
\label{fig2:pfcr}
\end{figure}

The proton number factorial cumulant ratios (normalized by $\kappa_1$) across 7 BES-II collider energies as a function of rapidity window size is presented in \Cref{fig2:pfcr}.
STAR data show negative two-proton and positive three-proton correlations, with amplitudes growing with rapidity window size, while fourth-order factorial cumulants remain near zero.
Near the critical point, when correlation length $\xi$ remains large enough, that the correlated range in rapidity exceeds the typical measurement window $\Delta y\ll \Delta y_\mathrm{corr}$, the factorial cumulants of event-by-event proton number distributions, are expected to exhibit a power-law dependence on the rapidity window of proton acceptance,  namely $\kappa_n\sim(\Delta y)^n$, or equivalently $\kappa_n/\kappa_1\sim(\Delta y)^{n-1}$ for normalized values~\cite{Ling:2015yau}.
In \Cref{fig2:pfcr}, for the most central (0 -- 5\%) collisions, power function fitting are performed to the STAR data.
All exponents from power-law fits to the data are lower than those expected near the critical region, suggesting the critical point may lie in a different energy domain. 
For example, the third-order fitted values, labeled in green, range from 0.8 to 1.7, all below the expected value of 2.

\subsection{Finite-Size Scaling with Susceptibility}

\begin{figure}[bh]
    \minipage[t]{0.48\linewidth}
    \centering
    \includegraphics[width=0.95\linewidth]{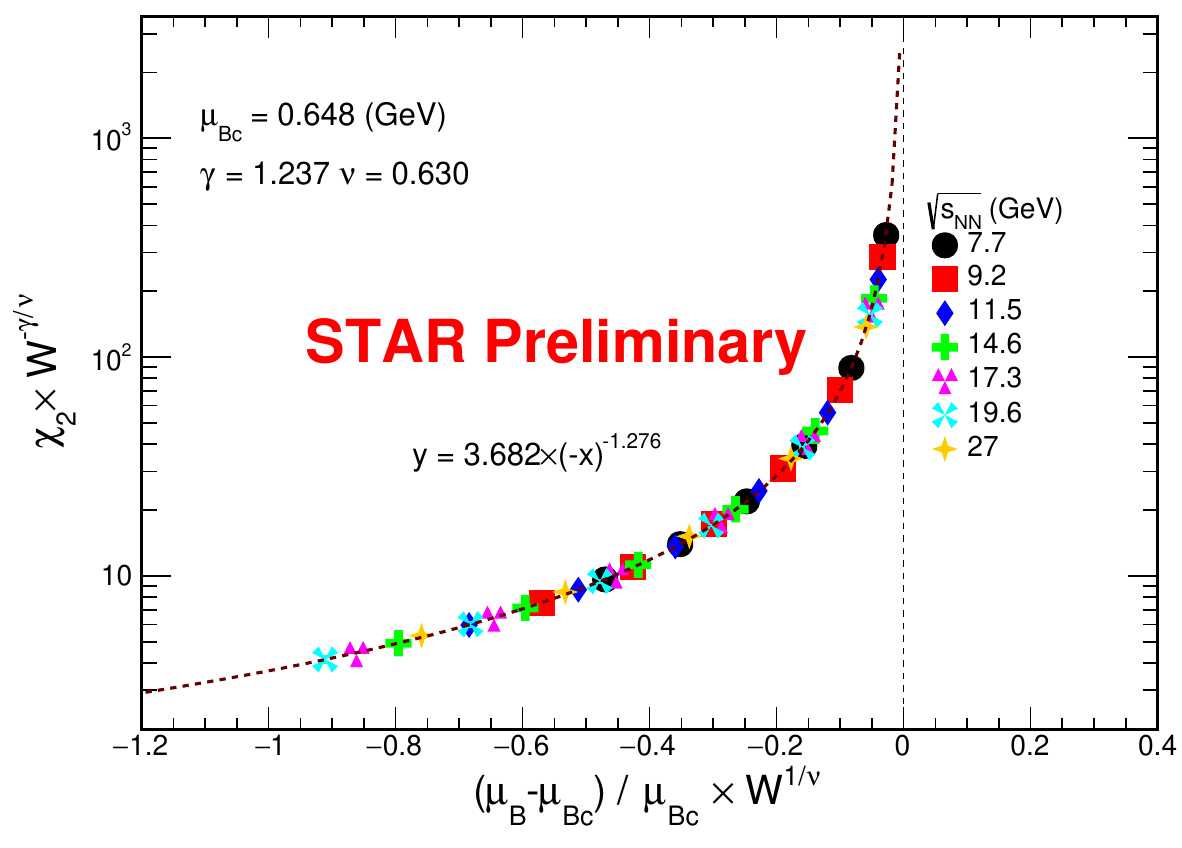}
    \caption{The finite-size scaling analysis of $\chi_2$ presents good scaling for data at $\sqrt{s_\mathrm{NN}}$ =7.7 -- 27 GeV. STAR data within various rapidity windows $W$ collapse onto the red dashed power-law curve well with critical baryon chemical potential $\mu_\mathrm{Bc}=648$ MeV.}
    \label{fig3:fss}
    \endminipage
    \hfill
    \minipage[t]{0.48\linewidth}
    \centering
    \includegraphics[width=0.95\linewidth]{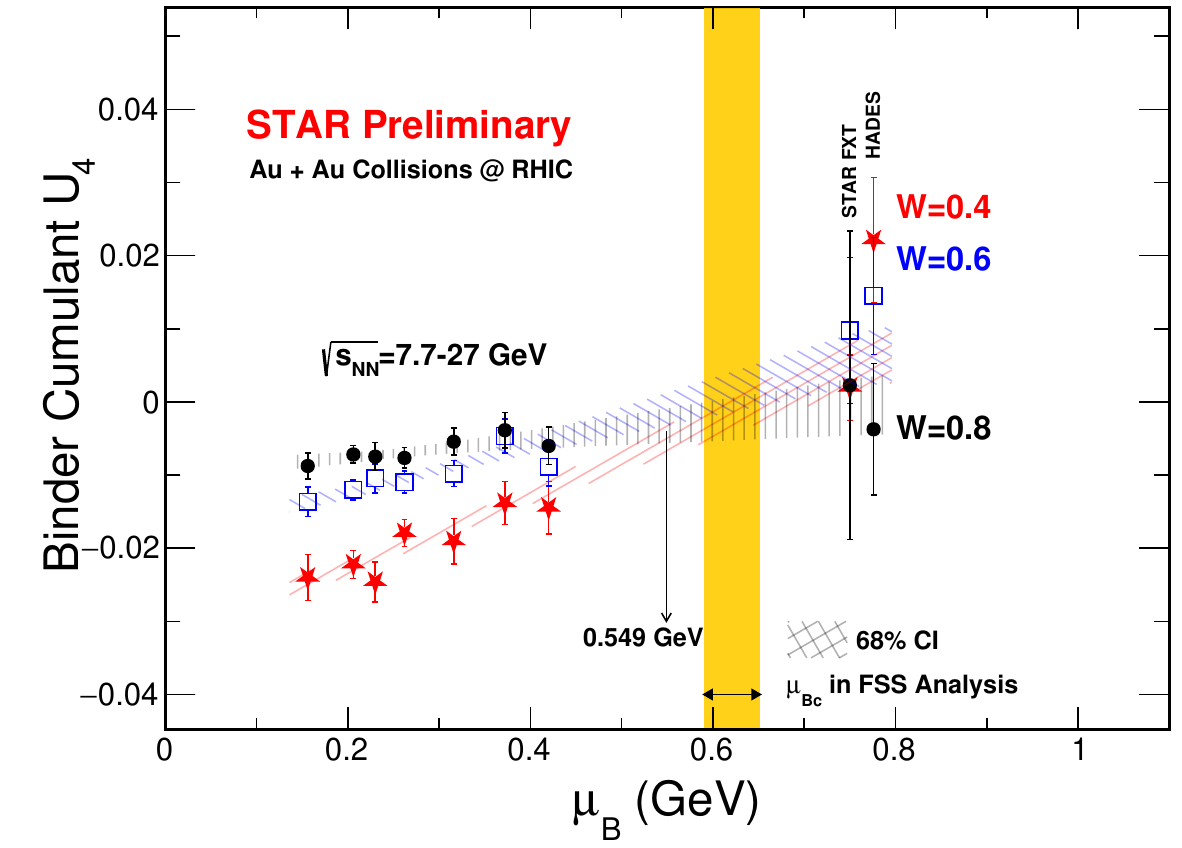}
    \caption{Binder cumulant as a function of $\mu_\mathrm{B}$. Markers represent different rapidity windows: $W=0.4$ (red stars), $W=0.6$ (blue squares), and $W=0.8$ (black dots). Linear fits yield 68\% confidence bands, with their overlap beginning around 549 MeV, consistent with the finite-size scaling estimate shown by the orange band.}
    \label{fig4:binder}
    \endminipage
\end{figure}

The susceptibility $\chi_n^\text{B,S,Q}$ can be connected to corresponding cumulants of the same order by $\chi_n^\text{B,S,Q}=C_n^\text{B,S,Q}/VT^3$.
When the correlation length becomes large, the system approaches scale invariance within a finite volume, making the sub-volume size the only relevant scale. 
This leads to the finite-size scaling (FSS) form for susceptibility, $\chi(L,t)=L^{\gamma/\nu}\Phi(tL^{1/\nu})$~\cite{Sorensen:2024mry}, where properly scaled data from different system sizes ($L$) collapse onto a universal curve ($\Phi$), allowing the critical point to be estimated by tuning its assumed location in $t$.
The susceptibility can be derived from measured net-proton cumulant and freeze-out parameters~\cite{Andronic:2017pug} via $\chi_n(W,\mu_\mathrm{fo})=C_n(W,\mu_\mathrm{fo})/(T_\mathrm{fo}^3W\mathrm{d}V_\mathrm{fo}/\mathrm{d}y)$,
where $W$ is the rapidity window size, $\mu_\mathrm{fo}$, $T_\mathrm{fo}$ and $\text{d}V_\mathrm{fo}/\text{d}y$ are thermal parameters: chemical potential, temperature, and volume per rapidity unit on the freeze-out curve. 
As shown in \Cref{fig3:fss}, using $t=(\mu_\mathrm{B}-\mu_\mathrm{Bc})/\mu_\mathrm{Bc}$ and the rapidity window size $W=\Delta y$, with $\gamma=1.237$ and $\nu=0.630$~\cite{sengers2009experimental}, we plot $\chi_2W^{-\gamma/\nu}$ versus $(\mu_\mathrm{B}-\mu_\mathrm{Bc})/\mu_\mathrm{Bc}\times W^{1/\nu}$. 
By scanning $\mu_\mathrm{Bc}$ and selecting the value with the smallest $\chi^2/\mathrm{ndf}$, we obtain $\mu_\mathrm{Bc}=648^{+4}_{-3}(\text{fit})^{+2}_{-58}(\text{sys.})$ MeV.
The fit uncertainty is determined using $\Delta\chi^2=1$, while the systematic uncertainty is evaluated by removing data points from the fit and by replacing the critical exponents with $\gamma=1$ and $\nu=0.5$.
Both fit and systematic uncertainties are rounded to integers.
This value is consistent with the results from BES-I in Ref.~\cite{Sorensen:2024mry} with significantly reduced fit uncertainty.

\subsection{Extrapolation with Binder Cumulant}

The Binder cumulant $U_4=-3C_4/C_2^2$ is related to both system size and correlation length~\cite{binder1981finite}.
We plot the Binder cumulant within different rapidity window $W$ as a function of $\mu_\mathrm{B}$ as shown in \Cref{fig4:binder}.
At the critical point, the finite system size becomes equivalently negligible compared to the infinite limit, making the Binder cumulant independent of system size. 
Consequently, data curves for different $W$ are expected to intersect at a common point, the critical point. 
The data to the right of the orange band come from STAR FXT at $\sqrt{s_\mathrm{NN}}=3$ GeV~\cite{STAR:2022etb} and HADES at 2.4 GeV~\cite{HADES:2020wpc}. 
For 3 GeV, $W$ is defined as $0-y_\mathrm{min}$, covering only negative rapidity, whereas for others $W$ is symmetric about zero.
On the left side, the markers show a clear ordering (black $>$ blue $>$ red) and increase with $\mu_\mathrm{B}$, with the growth rate reversed, leading to a change in order on the right. 
Linear fits to $W=0.4$ (red), $W=0.6$ (blue), and $W=0.8$ (black) yield 68\% confidence bands, whose overlap begins around 549 MeV. 
The expected intersection lies within this region, consistent with the finite-size scaling estimate.

\section{Summary}

This proceeding presents precise measurements of kinematic scans for (net-)proton number fluctuations in the STAR BES-II program, in Au+Au collisions covering collision energies from 7.7 -- 27 GeV.
The long-range effect near the critical point is explored through:
(1) power-law behavior of factorial cumulants,
(2) finite-size scaling of susceptibility across different rapidity windows, and
(3) Binder cumulant as a function of baryon chemical potential.
These studies point to a potential critical region in baryon chemical potential $\mu_\mathrm{B}$ in 550 -- 650 MeV.
This highlights the need for further investigations at high baryon density~\cite{Luo:2022mtp}.

\noindent \textbf{\textit{Acknowledgment}}: We thank the RHIC Operations Group and RCF at BNL. This work was supported by National Key Research and Development Program of China (No.2022YFA1604900), National Natural Science Foundation of China (NO. 12525509 and 12447102).

%
%
\bibliography{ref}

\end{document}